\documentclass[a4paper,11pt]{article}
\usepackage{pos}
\usepackage{float}

\newlength{\bibitemsep}
\newlength{\bibparskip}
\let\oldthebibliography\thebibliography
\renewcommand\thebibliography[1]{%
  \oldthebibliography{#1}%
  \setlength{\parskip}{\bibitemsep}%
  \setlength{\itemsep}{\bibparskip}%
}

\title{Overview of the EUSO-SPB2 Target of Opportunity program using the Cherenkov Telescope
}
 \ShortTitle{Overview of the EUSO-SPB2 Target of Opportunity program}

\author*[a]{Tobias Heibges}
\author[b]{Jonatan Posligua}
\author[a]{Hannah Wistrand}
\author[c,d]{Claire Gu\'epin}
\author[b]{Mary Hall Reno}
\author[e]{Tonia M. Venters}

\affiliation[a]{Colorado School of Mines, Department of Physics,\\
Golden, CO, USA}

\affiliation[b]{University of Iowa,Department of Physics and Astronomy,\\
  Iowa City, IA,  USA}

\affiliation[c]{University of Chicago, KICP,\\
Chicago, IL, USA}

\affiliation[d]{Laboratoire Univers et Particules de Montpellier,
\\
Montpellier, France}

\affiliation[e]{Goddard Space Flight Center,\\
Greenbelt, MD, USA}

\onbehalf{for the JEM-EUSO Collaboration}

\emailAdd{theibges@mines.edu}

\abstract{During the Extreme Universe Space Observatory on a Super Pressure Balloon 2 (EUSO-SPB2) mission, we planned Target of Opportunity (ToO) operations to follow up on possible sources of $\gtrsim 10 \, {\rm PeV}$ neutrinos.  The original plan before flight was to point the onboard Cherenkov Telescope (CT) to catch the source's path on the sky just below Earth's horizon. By using the Earth as a tau-neutrino to tau-lepton converter, the CT would then be able to look for optical extensive air shower signals induced by tau-lepton decays in the atmosphere. The CT had a field of view of $6.4^\circ$ vertical $\times$ $12.8^\circ$ horizontal. Possible neutrino source candidates include gamma ray bursts, tidal disruption events and other bursting or flaring sources. In addition, follow-ups of binary neutron star mergers would have been possible after the start of the O4 observation run from LIGO-Virgo-KAGRA. The resulting exposure is modeled using the NuSpaceSim framework in ToO mode. With the launch of the EUSO-SPB2 payload on the 13th May 2023, this summarizes the ToO program status and preliminary data, as available.}

\ConferenceLogo{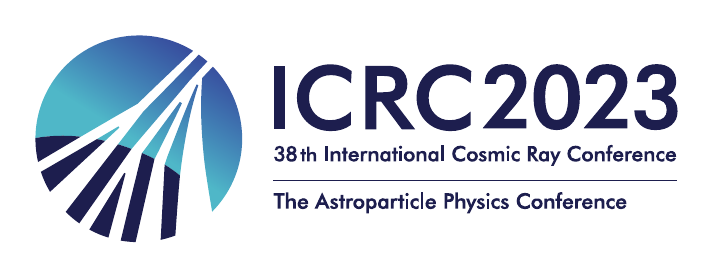}

\FullConference{%
The 38th International Cosmic Ray Conference (ICRC2023)\\
  26 July -- 3 August, 2023\\
  Nagoya, Japan}

\begin{document}
\maketitle

\section{Introduction}

The Target of Opportunity (ToO) program was devised as a way to collect interesting source candidates that produce neutrino-induced extensive air shower events and determine possible neutrino sources observed from the Extreme Universe Space Observatory on a Super Pressure Balloon 2 (EUSO-SPB2) \cite{2023ICRCEser,Eser:2021mbp}. This mission flew a setup of two telescopes, looking for very-high energy to ultra-high energy cosmic rays (UHECRs) and very-high energy neutrinos (VHE$\nu$) on a NASA super pressure balloon. The Fluorescence Telescope (FT) looked in nadir for fluorescence light from UHECR induced air showers of energies $>3\,$EeV \cite{Filippatos_2021}. The Cherenkov Telescope (CT) looked for Cherenkov light signatures from air showers induced either by cosmic rays of energies $\gtrsim 10\,$PeV or $\tau$-lepton ($\tau$) decays, which originate from $\tau$-neutrino ($\nu_\tau$) interactions inside the Earth's crust. 

\begin{figure}[b]
    \centering
    \includegraphics[width=\textwidth]{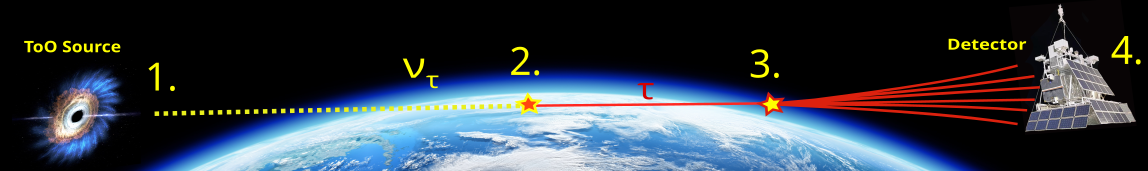}
    \caption{Illustration of the earth skimming method used in the ToO searches. 1. A neutrino source rises or sets below the horizon as viewed by the CT. 2. A neutrino from the source crosses through the Earth and possibly interacts 3. If a $\tau$ is created in this interaction, it could propagate and leave the Earth. 4. The $\tau$ decays in the atmosphere and produces an extensive air shower, which is observed by the CT.}
    \label{fig:ToO-method}
\end{figure}

\begin{figure}
    \centering
    \includegraphics[width=.5\textwidth]{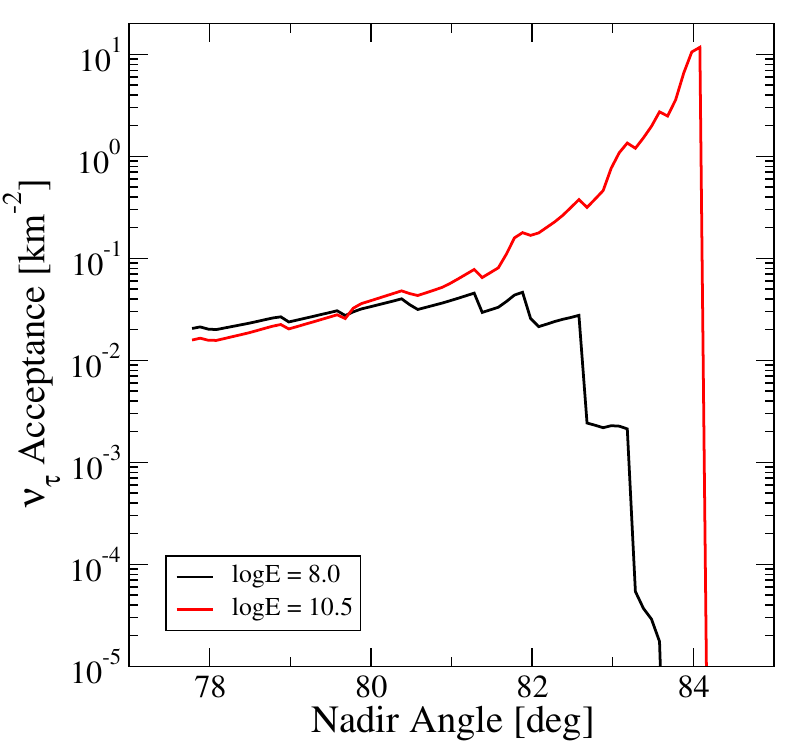}
    \caption{Sketch of the $\nu_\tau$ acceptance of the CT as a function of the nadir angle. It can be seen that higher energies have a larger acceptance closer to the limb ($84.2^\circ$ at a float altitude of 33\,km) than lower energies. Therefore, changing the tilt angle of the telescope also changes the sensitivity toward different energies.}
    \label{fig:enter-label}
\end{figure}

The ToO program focused on observations with the CT to look for $\nu_\tau$s that skim the Earth just below the horizon. This gives the $\nu_\tau$ a path-length through dense material inside the Earth to interact and produce a $\tau$. These $\tau$s have a probability to leave the Earth and produce extensive air showers detectable from EUSO-SPB2 \cite{Garg:2022ugd}. Since the neutrino cross section increases for increasing energies, there is a trade-off between the path-length through the Earth and what energy is expected to show maximum probability to produce a detectable air shower. A sketch of this process is shown in Figure \ref{fig:ToO-method}. Higher energy $\nu_\tau$s are expected to have a higher observation probability closer to the limb than their lower energy counterpart (see Figure \ref{fig:enter-label}). This method allows us to reach instantaneous sensitivities similar to ground based detectors. The goal of the ToO program was to identify and observe possible source candidates and calculate the times when their trajectories intercept the Earth near the horizon. Due to the duration of balloon flights, the most competitive limits are expected for transient sources. Therefore, transient sources are prioritized for follow-up observations. The exposure is calculated using the NuSpaceSim framework \cite{2023ICRCKrizmanic} in ToO mode. 
During flight, we produced a list of possible VHE$\nu$ sources such as blazar flares, tidal disruption events or gamma-ray bursts, observed by other observatories such as Fermi or Swift. Selecting the sources that pass within the CT's field of view, we produced tentative schedules for performing follow-up observations of these sources. 
We use the data collected during the flight of EUSO-SPB2 to look for a possible neutrino signal in excess over the background. The result of this search can be turned into a limit on the neutrino flux using a Feldman-Cousins 90\% confidence level upper limit \cite{Feldman:1997qc}. During the flight, measurements of the neutrinos with the telescope pointing below the Earth's limb have been performed and are currently under investigation to better understand the detector and search for possible events. So far no candidates have been found. 

\section{The Detector}

\begin{figure}[b]
    \centering
    \includegraphics[width=0.8\textwidth]{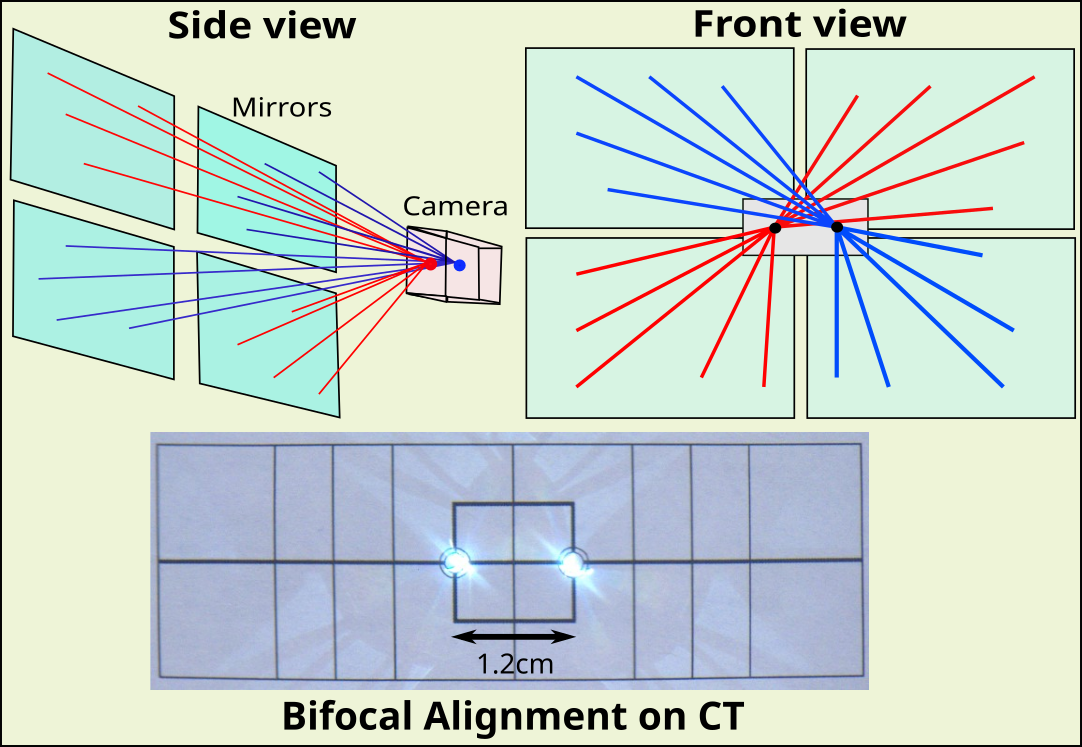}
    \caption{Sketch of the bifocal optics utilized in the CT. Diagonal pairs mirror segments (light blue boxes) are aligned to focus to two distinct spots (blue and red) on the camera (pink box) with the second one being offset by 1.2\,cm (one pixel in between). This is meant to reject direct cosmic ray hits as a background source.}
    \label{fig:bifocal_optics}
\end{figure}

The Cherenkov Telescope (CT) is a 1\,m diameter telescope, set up in a bifocal Schmidt optics configuration. There are four mirror segments which are aligned to create a bifocal image on the focal surface of the camera (see figure \ref{fig:bifocal_optics}). Parallel incident light focuses on two spots separated by $1.2\,\text{cm}$. The bifocal alignment of the mirrors is meant to reject direct cosmic hits and to reduce accidental triggers from background. The field of view for this detector is $6.4^\circ$ vertical $\times$ $12.8^\circ$ horizontal. 

The CT camera is a silicon photomultiplier camera with 512 pixels arranged in a $16\times 32$ pixel matrix. Each pixel has a size of $6\,\text{mm}\times 6\,\text{mm}$ and covers a field of view $0.4^\circ \times 0.4^\circ$. They are arranged in a $4\times 8$ matrices of $4\times 4$ pixels each. Each of these matrices is readout by two chipsets ("MUSIC-chips") that cover the left and right $2 \times 4$ pixels. The signal of the MUSIC-chips is digitized on a 10\,ns cadence. More information on the camera can be found in \cite{2023ICRCGazda}. 

The CT is attached to the main structure of EUSO-SPB2 (gondola) which hangs on a light weight NASA rotator. This rotator allows for $360^\circ$ rotation and pointing with $\pm 5^\circ$ precision. The front of the CT is attached to a linear stage which allows it to tilt up and down. This setup makes it possible to change the elevation angle of the optical axis of the CT from $3.5^\circ$ above horizontal to $-13^\circ$ below horizontal at a rate of $\sim 0.8^\circ/\text{min}$. This allows to look 
above the limb to test our sensitivity to cosmic ray induced extensive air showers' Cherenkov signals. Moreover, this allows to do follow up observations on neutrino sources below the limb.




\section{Source types and creating a schedule}

\begin{figure}[t]
    \centering
    \includegraphics[width=\textwidth]{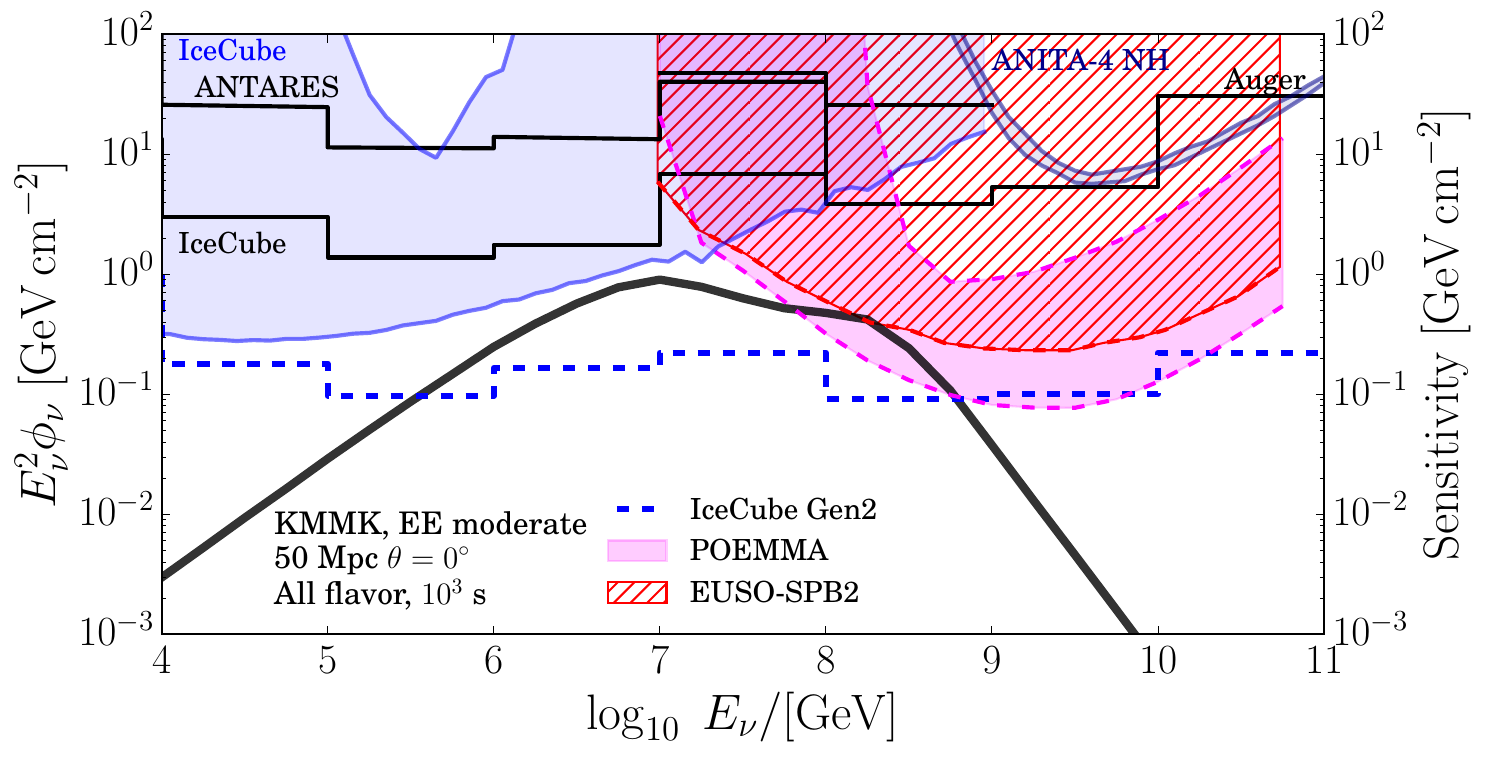}
    \caption{Projected short (1000 s) burst all-flavor neutrino plus antineutrino spectral fluence sensitivity for POEMMA and EUSO-SPB2. Also shown are 90\% confidence limits from ANTARES, IceCube and Auger in a $\pm$500\,s window around the gravitational wave event from GW170817 adapted from \cite{Ackermann:2022rqc}, with POEMMA \cite{Venters:2019xwi} and EUSO-SPB2 \cite{Venters_2021ICRC}. The projected sensitivities of the CT over a time period of 1000\,s are expected to generate sensitivities similar to current ground-based detectors. For reference, the extended emission short gamma ray burst flux by Kimura et al. \cite{Kimura_2017}  (on-axis viewing ($\theta = 0^\circ$) from a source 50\,Mpc distant) is shown with the solid black curve.}
    \label{fig:spb2-sensitivity}
\end{figure}

Our integrated exposure is limited by night time operations only, short observation periods on any particular source, and limited flight time for a balloon mission. However, the expected instantaneous acceptance toward VHE$\nu$ is expected to be comparable with other ground based detectors (see figure~\ref{fig:spb2-sensitivity}). Therefore, VHE$\nu$ searches on EUSO-SPB2 are especially powerful for transient sources. These include binary neutron star mergers, blazar flares, tidal disruption events, gamma-ray bursts, core collapse supernovae, and others \cite{Guepin:2022qpl}. On top of transient sources, known steady sources of interest are added to the possible source list. 
More on these can be found in \cite{2023ICRCWistrand}.

A software package has been implemented to automate the process of collecting possible sources and extracting needed information from the alerts. It is currently undergoing further development and in the near future is planned to be made publicly available. The framework for this package is detailed in the following subsections and in \cite{2023ICRCPosligua}.

\textbf{Alert collection and database} - The alerts are collected from the Global Coordinate Network (GCN), Transient Name Server (TSN) and Astronomer's Telegrams (ATels). For both GCN and TNS, automatic listening services have been implemented that extract the coordinates and other helpful information from the event notices and written into a database. The GCN listening service is updated in real time, whereas  the TNS listener fetches new alerts once every hour. Since ATels are not machine readable, they have to be added manually to the database.

\textbf{Observability calculation} - Each of the sources in the database is propagated over the sky for a time period of 36\,h, and a list of observable sources is computed, with their locations and observation times. Sources are considered observable with the CT when they cross the horizon during the night, with the Moon set or a low Moon illumination. A special algorithm has been implemented for extended sources such as gravitational wave events. By splitting the region of interest into areas of equal solid angle regions and weighting each section by the probability of the source being in the section, an effective location using the probability weighted time is calculated   \cite{2023ICRCPosligua}.

\textbf{Source prioritization and selection} - From the list of all observable sources a subset needs to be selected for any given night, due to the large number of observable sources and the limited observation time. A prioritization scheme for the sources is implemented, based on flux predictions from models, distance of the source from Earth and relative occurrence rate in the local universe \citep[e.g.,][]{Venters:2019xwi, Guepin:2022qpl}. Using this as a guide, sources are classified in different priority tiers. From higher to lower priority: 1/ galactic transients (rare events: supernova, binary neutron star merger, tidal disruption events), 2/ extra-galactic binary neutron star mergers, IceCube gold or bronze events \cite{blaufuss2019generation}, 3/ tidal disruption events, 4/ blazar flares, 5/ gamma ray bursts, 6/ supernovae, 7/ steady sources. More on the rationale of this scheme can be found in \cite{2023ICRCWistrand}. After the priorities are assigned for each source, a subset needs to be selected for follow up observations for each night. 
An algorithm computes the observation schedule for each night. It picks all the sources from the tiers 1 and 2. Then it picks one source from each priority tier $>2$ and then moves on to the next until it runs out of tiers or sources. If it runs out of tiers, it restarts from the highest tier. This ensures a good mix between the different source types for the mid tiers (3-6), for which the comparison between the flux predictions from different models is uncertain. For the selection within a tier, a set of different methods has been implemented, which can be chosen by the user, for instance: select a random source from each tier, select the most recent source from each tier, select the source with the longest observation time from each tier, select sources that have been previously observed.

\textbf{ToO mode in NuSpaceSim} - As a future extension of the ToO software, the NuSpaceSim framework will be used to estimate the acceptance for a source more quantitatively \cite{Venters_2021ICRC}. In the current version, the time taken by a source to cross the CT's field of view is used as a proxy for the acceptance.
NuSpaceSim is a simulation framework to calculate the acceptance for VHE$\nu$ sources as seen from a balloon or satellite mission. It was originally designed to calculate the acceptance of a diffuse neutrino flux and is currently being modified to incorporate a ToO-mode, where the acceptance can be calculated for point sources. This is done by throwing random trajectories within the projection of the Cherenkov cone from the detector onto the surface of the Earth that coincide with the direction of the source in the sky. For each of these trajectories, the interaction probability and the $\tau-$exit probability is calculated. Based on this, an air shower signal can be estimated and averaged over all possible trajectories. This can be used to estimate the acceptance of a detector for any given point source in the sky. At this time, the ToO software is still under development and final testing, and it will be accessible in the near future.

\section{Backgrounds}
There are several potential sources of backgrounds to VHE$\nu$ observations with the CT. Some of the main sources of background that were identified so far are listed below.

\textbf{Direct cosmic ray hits} - While the rate of direct cosmic ray hits is still under some investigation, the bifocal optics reduces the rate to random coincidence to be within the two bifocal pixels and the same digitization bin, which is expected to be very small.

\textbf{Refracted air showers} - The density of the atmosphere as a function of altitude leads to an altitude dependent index of refraction, and therefore, refraction of Cherenkov light from extensive air showers. More in depth analysis of this effect show that the light is bent away from the detector. Refracted air shower events are therefore no significant source of background to ToO searches.

\textbf{Reflected air showers} - Reflections of the Cherenkov light from an EAS from clouds or the ocean surface could then mimic neutrino events coming from below the limb. The signature and rate of these events is still under investigation.

\textbf{Deflected muon induced subshower} - Over the long pathlength through the Earth's magnetic field, muons could be significantly deflected from the original shower axis and create short EAS like signals in the CT. Further analysis is performed in \cite{2023ICRCFuehne}.

\textbf{Night sky background} - Stray light can produces light flashes that can mimic bifocal signals in the CT. These events are expected and the instrument thresholds get set so that the background rate is $\sim 10\,{\rm Hz}$ in the detector. 

\textbf{Electrical noise} - There are also different types of background electronic sources, such as accidental triggers and cross talk. For more information on these background sources, see \cite{2023ICRCGazda}.

\section{In flight performance}
\begin{figure}[t]
    \centering
    \includegraphics[width=.49\textwidth]{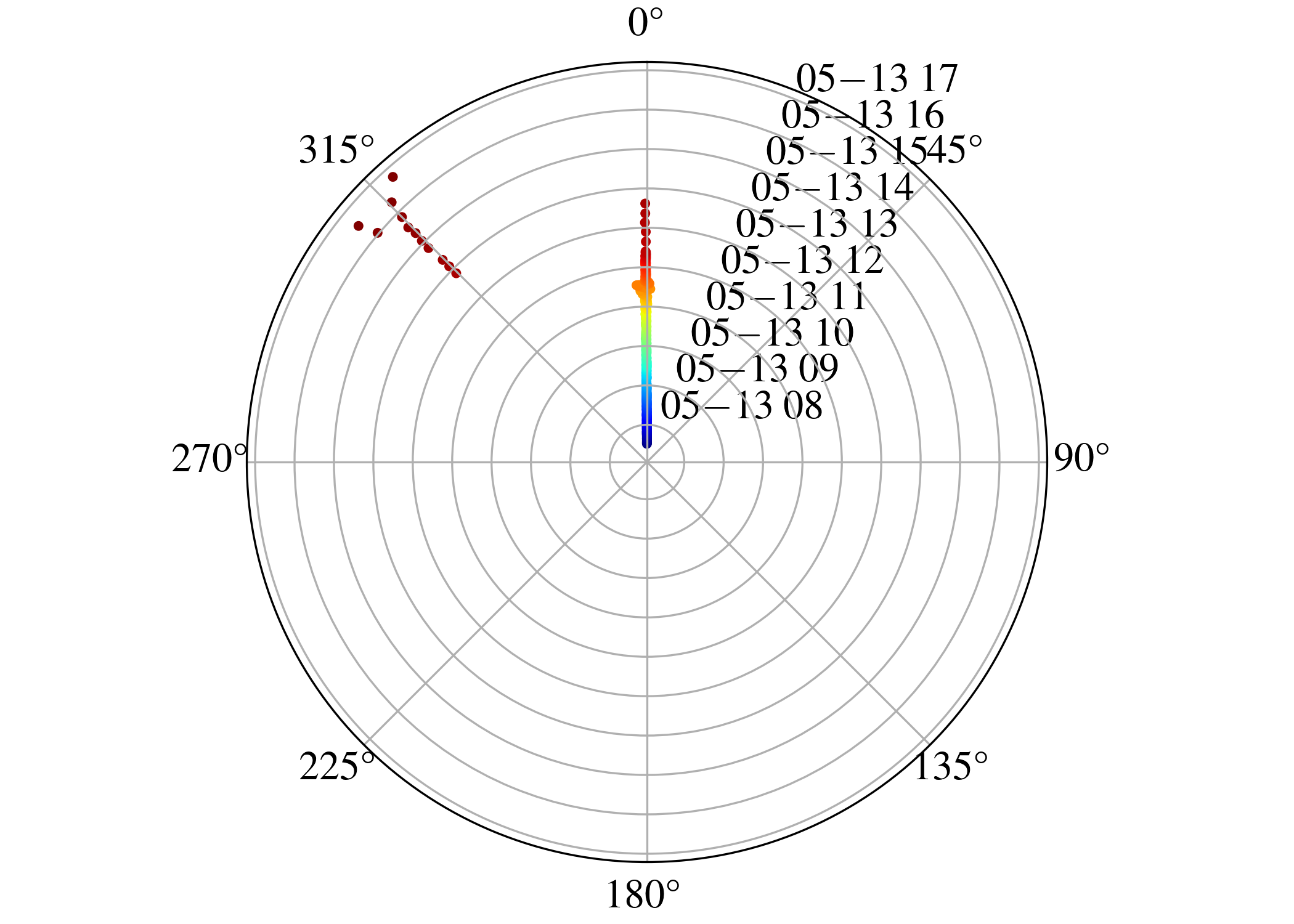}
    \includegraphics[width=.49\textwidth]{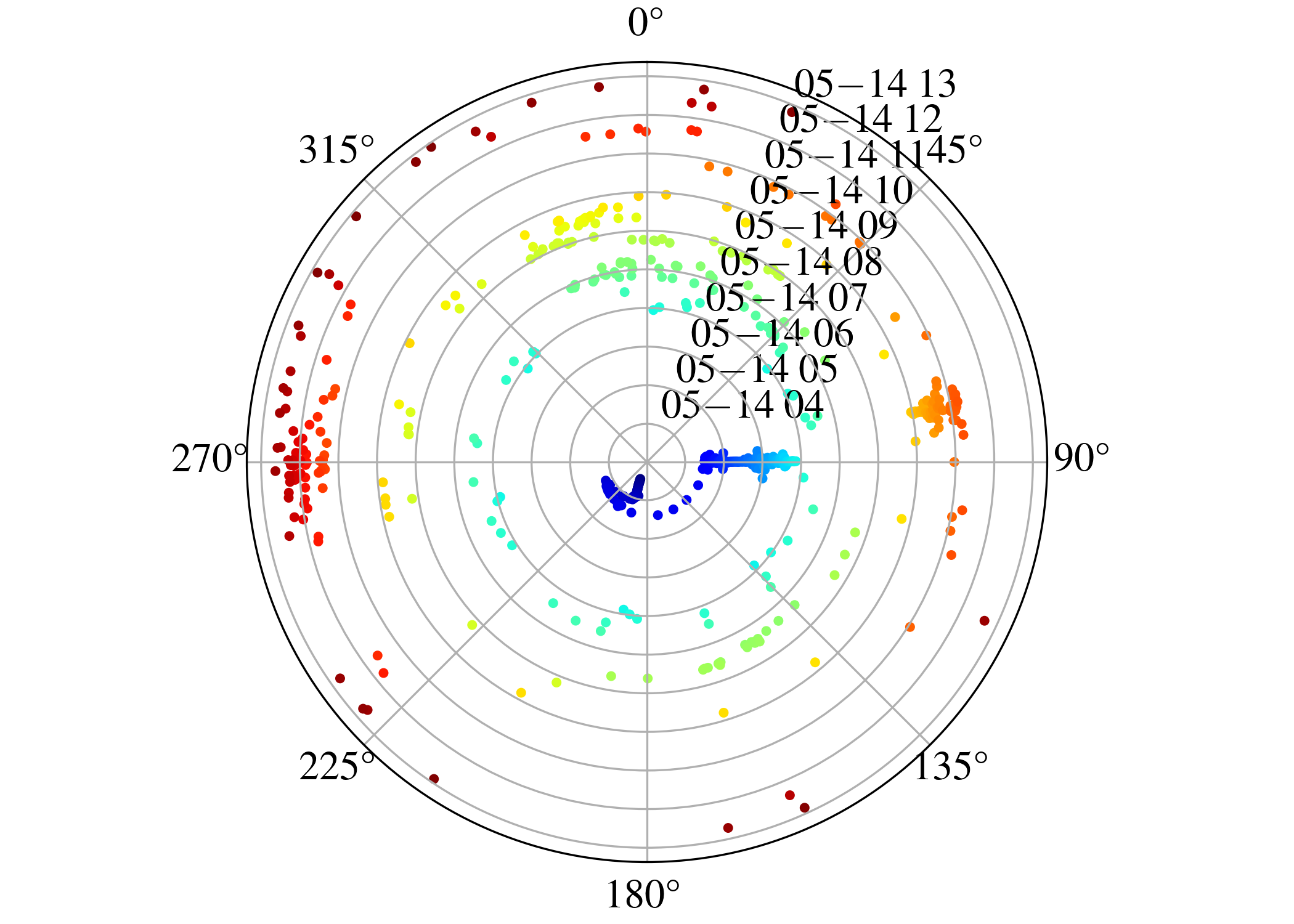}
    \caption{Pointing directions of the CT for night 1 of data taking (left) with nominal operation and stable pointing over the whole night and night 2 (right) during descent and very unstable pointing direction due to wind and the descent.}
    \label{fig:pointing}
\end{figure}

The EUSO-SPB2 mission was launched from Wanaka on the 13th of May 2023 at 00:02 UTC and unfortunately was terminated over the Pacific Ocean on the 15th of May at 12:55  UTC after only 1 day 12 hours and 53 minutes. This time period included two periods of astronomical darkness during which observations with both telescopes were possible, which showed that all instruments onboard were operational and ready for many hours of data-taking. The  first night on May 13th was dedicated to commissioning the instrument and understanding the observation conditions. This included turning on the CT camera and HV on with the shutter doors closed and open as well as testing the shutter and tilt system. We also tested the operation of the rotator responsible for the azimuth pointing, which can be seen in the left panel of figure~\ref{fig:pointing}, indicating that systems are operational and capable of performing ToO follow-up observations on consecutive nights. More on the commissioning of the CT camera can be found in \cite{2023ICRCGazda, 2023ICRCRomero}.

After we learned that the balloon had a leak, the second night was focused on data-taking and maximizing the value extracted from the flight. This included two periods of looking for neutrino signals from below the limb and about $45\,$min of observations of cosmic-ray signals conducted above the limb. During the time period we were pointing above the limb, several bifocal cosmic-ray candidate events have been observed, showing that the camera was functional and capable of detecting such signals \cite{2023ICRCGazda, Cummings:2020ycz, 2023ICRCCummings}. An attempt to follow up a possible ToO was made. However, at the time the balloon had dropped far enough that the friction in the air was too high for the rotator to overcome. This can be seen in the right panel of figure~\ref{fig:pointing}, which shows several time periods during which the balloon was spinning uncontrollably, and during the times when the pointing direction could be stabilized, the spread of pointing directions is significantly bigger than during the first night, shown in the left panel. Moreover, as the balloon was dropping in altitude, the direction of the horizon also changed leading to a longer pathlength for possible ToO observation for a fixed telescope elevation angle, which in turn lowered the sensitivity to higher energy showers. Based on the recovered pointing directions of the balloon, a list of `accidental' ToO observations  over the course of the second night of the flight is under construction. Also, data searches for possible neutrino candidates in the over the $\sim 30000$ recorded bifocal triggers over the whole flight and analysis of the backgrounds are ongoing with no candidates found yet.

\section{Conclusion}
The EUSO-SPB2 mission  placed a Cherenkov detector in a near-space configuration on a super pressure balloon. We showed that the instrument was fully functional and able to observe EAS signatures when pointed above the Earth's horizon. Furthermore, we demonstrated our ability to tilt and rotate the CT that allow for follow-up observations of ToO candidates. The ability to tilt above and below the horizon also allows for a built-in test of the instrument to show its functionality during neutrino searches. 
During our flight, we generated a list of possible neutrino sources, tilted the telescope below the limb, and performed a series of background observations. There is ongoing analysis regarding possible accidental observations of ToO sources during the descent of the balloon. While the abbreviated flight of EUSO-SPB2 cut our observation time and data collection very short, the framework developed for this project can be used independently and applied to other projects. Although this may not have been the extensive data collection that we had hoped for, the basis for this project will continue to be refined and adapted to future missions.\\
\small{\noindent{\bf Acknowledgements} -- The authors acknowledge the support by NASA awards 11-APRA-0058, 16-APROBES16-0023, 17-APRA17-0066, NNX17AJ82G, NNX13AH54G, 80NSSC18K0246, 80NSSC18K0473,80NSSC19K0626, 80NSSC18K0464, 80NSSC22K1488, 80NSSC19K0627 and 80NSSC22K0426, the French space agency CNES, National Science Centre in Poland grant n. 2017/27/B/ST9/02162, and by ASI-INFN agreement n. 2021-8-HH.0 and its amendments. This research used resources of the US National Energy Research Scientific Computing Center (NERSC), the DOE Science User Facility operated under Contract No. DE-AC02-05CH11231. We acknowledge the NASA BPO and CSBF staffs for their extensive support. We also acknowledge the invaluable contributions of the administrative and technical staffs at our home institutions.}

{\small
\bibliographystyle{JHEP-nt}
\bibliography{bibliography}}

\clearpage

\newpage
{\Large\bf Full Authors list: The JEM-EUSO Collaboration\\}

\begin{sloppypar}
{\small \noindent
S.~Abe$^{ff}$, 
J.H.~Adams Jr.$^{ld}$, 
D.~Allard$^{cb}$,
P.~Alldredge$^{ld}$,
R.~Aloisio$^{ep}$,
L.~Anchordoqui$^{le}$,
A.~Anzalone$^{ed,eh}$, 
E.~Arnone$^{ek,el}$,
M.~Bagheri$^{lh}$,
B.~Baret$^{cb}$
% ...
}
\end{sloppypar}
\vspace*{.3cm}

{ \footnotesize
\noindent
% ...
$^{cb}$ Universit\'e de Paris, CNRS, AstroParticule et Cosmologie, F-75013 Paris, France\\
% ...
$^{ed}$ Istituto Nazionale di Fisica Nucleare - Sezione di Catania, Italy\\
% ...
$^{eh}$ INAF - Istituto di Astrofisica Spaziale e Fisica Cosmica di Palermo, Italy\\
% ...
$^{ek}$ Istituto Nazionale di Fisica Nucleare - Sezione di Torino, Italy\\
$^{el}$ Dipartimento di Fisica, Universit\`a di Torino, Italy\\
% ...
$^{ep}$ Gran Sasso Science Institute, L'Aquila, Italy\\
% ...
$^{ff}$ Nihon University Chiyoda, Tokyo, Japan\\ 
% ...
$^{ld}$ University of Alabama in Huntsville, Huntsville, AL, USA\\
$^{le}$ Lehman College, City University of New York (CUNY), NY, USA\\
% ...
$^{lh}$ Georgia Institute of Technology, USA\\
% ...
}

\end{document}